\documentclass[11pt, letterpaper]{article}

\usepackage{amsmath}
\usepackage{amssymb}
\usepackage{amsthm}
\usepackage{graphicx}
\usepackage{hyperref}
\usepackage{url}
\usepackage{algorithm}
\usepackage{algorithmic}
\usepackage{caption}
\usepackage{subcaption}
\usepackage{booktabs}
\usepackage{authblk}
\usepackage{appendix}
\usepackage[utf8]{inputenc}
\usepackage[T1]{fontenc}
\usepackage[english]{babel}
\usepackage{geometry}
\usepackage{xcolor}
\usepackage{fancyhdr}
\usepackage{microtype}
\usepackage{parskip}
\usepackage{tabularx}
\usepackage{enumitem}

\usepackage{cite}
\hypersetup{
    colorlinks=true,
    linkcolor=blue,
    filecolor=magenta,      
    urlcolor=cyan,
    citecolor=red,
}

\definecolor{headercolor}{RGB}{0, 50, 100}
\newcommand*\samethanks[1][\value{footnote}]{\footnotemark[#1]}

\title{RadOnc-GPT: A Large Language Model for Radiation Oncology}

\author[1]{Zhengliang Liu, MS \thanks{Co-first authors.}}
\author[2]{Peilong Wang, PhD \samethanks}
\author[1]{Yiwei Li, MS \samethanks}
\author[2]{Jason Holmes, PhD}
\author[1]{Peng Shu, MS}
\author[2]{Lian Zhang, PhD}
\author[3]{Chenbin Liu, PhD}
\author[1]{Ninghao Liu, PhD}
\author[4]{Dajiang Zhu, PhD}
\author[5]{Xiang Li, PhD}
\author[5]{Quanzheng Li, PhD}
\author[2]{Samir H. Patel, MD}
\author[2]{Terence T. Sio, MD, MS}
\author[1]{Tianming Liu, PhD}
\author[2]{Wei Liu, PhD}

\affil[1]{School of Computing, University of Georgia}
\affil[2]{Department of Radiation Oncology, Mayo Clinic}
\affil[3]{Department of Radiation Oncology, Cancer Hospital Chinese Academy of Medical Sciences Shenzhen Center}
\affil[4]{Department of Computer Science and Engineering, University of Texas at Arlington}
\affil[5]{Department of Radiology, Massachusetts General Hospital and Harvard Medical School}

\date{}

\begin{document}

\maketitle

\begin{abstract}
This paper presents RadOnc-GPT, a large language model specialized for radiation oncology through advanced tuning methods. RadOnc-GPT was finetuned on a large dataset of radiation oncology patient records from the Mayo Clinic in Arizona. The model employs instruction tuning on three key tasks - generating radiotherapy treatment regimens, determining optimal radiation modalities, and providing diagnostic descriptions/ICD codes based on patient diagnostic details. Evaluations conducted by comparing RadOnc-GPT outputs to general large language model outputs showed higher ROUGE scores in these three tasks. The study demonstrated the potential of using large language models fine-tuned using domain-specific knowledge like RadOnc-GPT to achieve transformational capabilities in highly specialized healthcare fields such as radiation oncology. However, our model's clinical relevance requires confirmation, and it specializes in only the aforementioned three specific tasks and lacks broader applicability. Furthermore, its evaluation through ROUGE scores might not reflect the true semantic and clinical accuracy — challenges we intend to address in future research. 

\end{abstract}

\section{Introduction}

Large Language Models (LLMs) has come through remarkable development \cite{wang2023review,liu2023evaluating,liu2023summary,holmes2023evaluating}, The development in LLMs has spurred achievements in proposing and developing large foundation models in computer vision aspect \cite{dai2023hierarchical,wang2023review,dai2023samaug,zhang2023segment,li2023artificial,zhang2023beam}. 
While Large Language Models (LLMs) such as ChatGPT and GPT-4 have displayed remarkable capacities in natural language processing (NLP) of many areas \cite{guan2023cohortgpt,liu2023pharmacygpt,liu2023radiology,wu2023exploring,ma2023impressiongpt,chang2023meta,dai2023ad,cai2023exploring,qiang2023deep,rezayi2023exploring}, their direct application in niche areas like healthcare has posed challenges. 

Specifically, in radiation oncology, a sector demanding utmost precision, the generic nature of mainstream LLMs like ChatGPT or GPT-4 falls short. Distinct from other clinical practices, radiation oncology presents two salient features: a high level of complexity and stringent requirements for precision \cite{dawson2007advances, xing2006overview}. This involves a chain of activities such as initial consultation, multi-modal imaging-based simulation, treatment planning, quality assurance, plan execution, and patient follow-up \cite{unkelbach2018robust, li2015robust, cao2012uncertainty,an2017robust, li2012dynamically, shen2016efficient, liu2023artificial}. These steps incorporate a dynamic interplay between imaging data and text-based clinical information, continually updated and accurately communicated to inform the subsequent phase of radiation treatment. Ultimately, this intricate process aims to ensure optimal implementation of radiation therapy. This process is traditionally time-consuming, reliant on manual analysis of vast amounts of unstructured clinical data, and susceptible to variations in human interpretation. Efficient tools for language-involved processing can significantly enhance each phase of radiation therapy, and potentially improve treatment outcomes. This is especially true for advanced radiation therapy techniques requiring even greater complexity and precision, such as Intensity-Modulated Proton Therapy (IMPT) \cite{schild2014proton,younkin2018multiple, liu2013ptv, shan2022virtual, feng2022gpu}. The necessity arises for a model that assimilates clinical domain knowledge with the conciseness and specificity inherent in radiation oncology. Furthermore, medical institutes and healthcare practitioners need to have their localized LLMs because of the privacy regulations for patient health information (PHI). Besides, the collaboration of NLP and radiation oncology still remains a underlooked field which even has enormous amount of available text data. Some previous works have already expanded in medical domain\cite{eggmann2023implications,haupt2023ai,zhao2023brain}. In detail, fine-tuned LLMs can be used to generate radiotherapy treatment regimens and other medical reports\cite{liao2023differentiate,dai2023chataug,liu2023deid, ma2023impressiongpt}, determine radiation treatment modality such as proton versus photon therapy, predict diagnosis description or ICD code based on patient details. These models all indicate the potential capabilities of LLMs in radiation oncology. By dealing with clinical notes and other patient's medical diagnosis and images, it is able to construct a conversational model for clinicians and patients about radiation oncology and oncology in general. However, there still leaves a blank for radiation oncology to embrace such a comprehensive and powerful model.

In order to maximizing the potential of clinical notes in radiation oncology, we put forward a pioneering solution called RadOnc-GPT. RadOnc-GPT addresses above-mentioned issues by offering a specialized LLM tailored to generate radiation oncology impressions from clinical data, boasting improved clarity, specificity, and clinical relevance.

\section{Related work}

\subsection{Large Language Models (LLMs)}
The NLP landscape has seen a significant shift with the rise of LLMs such as GPT-4\cite{openai2023gpt}, PaLM\cite{chowdhery2022palm}. These models, characterized by their few-shot and zero-shot learning capabilities, mark a departure from the conventional pre-training and fine-tuning methodologies including BERT \cite{devlin2018bert}, GPT \cite{radford2018improving}, GPT-2 \cite{radford2019language}, and their variants \cite{liu2019roberta,liao2023mask,zhou2023comprehensive}. Moreover, models like Alpaca \cite{stanfordStanfordCRFM} and Dolly \cite{databricksFreeDolly} illustrate the growing trend towards instruction-based LLMs. These LLMs have shown great potentials in various domains and have profoundly affected the development of all walks of life. 

\subsection{Domain-Specific Language Models (DSLMs)}
For original LLMs, because they are trained on a wide range of general data, these LLMs have acquired very strong reasoning capabilities and knowledge reserves. However, when these models face questions with more professional orientation, the answers often fail to answer the questions and cannot grasp the answers. focus. More and more DSLMs are being trained for this problem. DSLMs, like AgriBERT \cite{rezayi2022agribert} and SciEdBERT \cite{liu2023context}, signify the evolution of models tailored to serve specialized sectors, which aim to outperform other models in related agriculture and education works. In the healthcare arena, ClinicalRadioBERT \cite{rezayi2022clinicalradiobert} emerges as a prime example, dedicated to radiation oncology. Such models underscore the increased relevance and flexibility of domain-focused LLMs. In public health domain, automatic tools like AD-AutoGPT \cite{dai2023ad} greatly increase the efficiency of news collection for researchers. However, DSLMs still have a lot of potential to be explored. There are many fields that provide application scenarios that are more suitable for LLMs, and a large amount of data suitable for fine-tuning the model has been accumulated. If it can be fully utilized, it can revolutionize more industries.

\section{Data \& Methodology}

RadOnc-GPT was trained using a vast dataset from radiation oncology, applying advanced tuning methods to craft radiation oncology impressions. Evaluations conducted by oncologists revealed RadOnc-GPT's superiority over traditional LLMs in terms of clarity, specificity, and clinical relevance in the impressions formulated.

\begin{figure}
    \centering
    \includegraphics[scale=1.0]{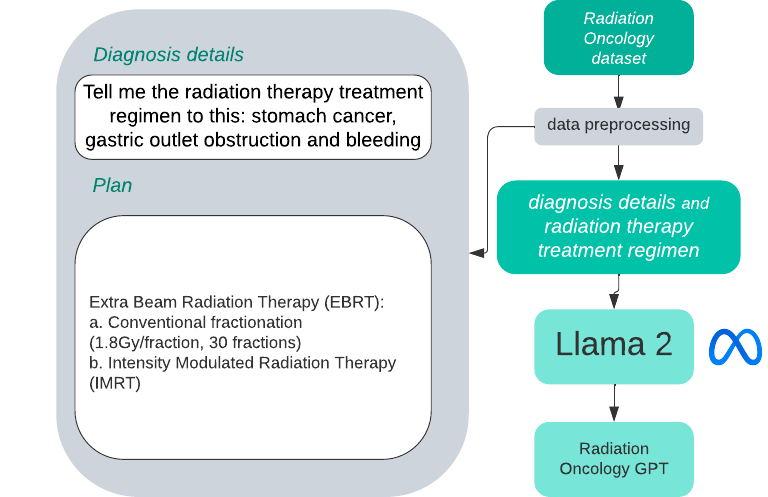}
    \caption{The overall framework of RadOnc-GPT.}
    \label{fig:framework}
\end{figure}

\subsection{Data extraction}
We utilized a python script to connect to the database of Aria ver.15.6 (Varian Medical Systems, Palo Alto, CA), which stores patient data on an SQL database running locally at the Department of Radiation Oncology of Mayo Clinic in Arizona. The python script extracted detailed patient diagnosis information. The python script is composed of an SQL query and post-processing of the data. The retrieved data encompasses the patient ID, ICD codes\footnote{ICD stands for “International Classification of Diseases,” ICD code is a unique alphanumeric code used to represent specific diseases, medical conditions, and health-related issues.}, onset dates, disease stages, and descriptions from relevant database tables. The post-processing step involves transforming ICD code indices into their standard textual notations (e.g., "ICD-10-CM" or "ICD-9"), prefacing diagnostic details\footnote{The "diagnostic details" provides a summary of the patient's disease history, current status, and the treatment plan prescribed by the physician. } with the descriptor "diagnosis details", and ordering the data by onset date and patient ID. Finally, the dataset was saved as a CSV file.

In total, we extracted 15,724 patient diagnostic cases from the cancer patients treated at the Department of Radiation Oncology of Mayo Clinic in Arizona. In an initial step, we selected cases where each patient had only one diagnostic record (unique cases) to simplify the fine-tuning process. This resulted in 9,799 cases.

\subsection{Data curation}
Data curation plays a crucial role in the fine-tuning and validation of our model. The ICD codes were well-documented within the diagnostic notes, making them directly usable without preprocessing. However, to separate the patient's disease history and status from the treatment plan, we performed a manual process. First, we used a simple keyword filter, primarily searching for the term "Plan" or similar keywords. Subsequently, we manually corrected and refined the separation. These steps successfully isolated 3,282 cases where the disease history and status were distinctly separated from the treatment plan. Additionally, we removed physician names from the treatment plans during this manual correction process to enhance the robustness of our fine-tuning database.

To further enhance the dataset, we manually labeled the radiation therapy treatment type for the selected 3,282 cases. These labels included terms such as "proton", "photon", "electron", "photon+brachy", etc. Both the preprocessed data and the labeled data underwent rigorous quality control checks.

After preprocessing and the labelling, we excluded patient ID, patient name, and onset date from the dataset to protect patient privacy. This refined data is now stored in a standalone database and ready for use in or adapted for our upcoming LLM fine-tuning tasks.

\subsection{Instruction Tuning}

\subsubsection{Instruction Tuning's Principle and Objective}

Instruction tuning is the crucial part of the RadOnc-GPT. It is the necessary step to make LLMs adapt to the given domain and answer the question in domain-specific formats. This technique requires the pairs of domain-specific instructions and desired outputs. The purpose of this approach is to synchronize the model with domain-specific tasks, improve the manageability of the model, and facilitate quick adaptation to specific domains. All these are achieved without compromising computational efficiency.

In the domain of Radiation Oncology, the majority of existing LLMs are unable to effectively address the questions pertinent to this field due to the high professional threshold and the confidentiality of patient data. Even though some LLMs can provide relevant background information, there remains a significant discrepancy in both format and content when compared to clinical reports. However, the reality is that a vast number of clinical notes are available for utilization, indicating a substantial potential for instruction fine-tuning within the realm of Radiation Oncology.

\subsubsection{Knowledge Acquisition}
LLAMA2 is a powerful large language model developed by Meta AI, which demonstrates excellent performance in common sense reasoning and knowledge acquisition. Compared to LLAMA, LLAMA2 used GQA to dramatically improve inference throughput. LLAMA2 trained two independent reward models, one optimized for helpfulness and the other optimized for safety.The performance improvement of the LLAMA2-chat model was proved in terms of reward model accuracy. Thus, the models obtained by LLAMA2 fine-tuning are among the best in various tasks. RadOnc-GPT, utilizing LLAMA2 as its foundational model, undergoes instruction fine-tuning across three fundamental tasks, thereby generating distinct model parameters to address each specific task. These tasks encompass three tasks which are:
\begin{itemize}
    \item \textbf{The generation of radiotherapy treatment regimens based on patient details:} The treatment regimen is a discipline that intertwines the domains of physics, biology, surgery, and medicine, aiming to maximize the therapeutic ratio of cancer treatment. The role of a radiation oncologist in this multidisciplinary arena is intricate and crucial, given that the decisions made will have long-lasting impacts on a patient's quality of life and overall survival. Various key considerations play a role in devising an effective treatment regimen. Initial steps involve an exhaustive diagnostic assessment that includes imaging studies such as CT scans, MRIs, or PET scans and may also include biopsy results for histopathological confirmation. The tumor staging based on TNM classification (Tumor, Nodes, Metastasis) is critical for selecting appropriate treatment options. The radiation oncologist must decide the objectives of the treatment regimen. Radiation therapy is commonly part of a broader treatment landscape that might include surgery, chemotherapy, or immunotherapy. Deciding, scheduling and sequencing the combination of different treatment procedures within the regimen must be harmonized for the best therapeutic outcome. Consideration of the patient’s overall health, existing comorbidities, and ability to tolerate treatment can affect decisions about regimen intensity, duration, and suitability for combining with different treatment procedures. After the initiation of therapy, regular follow-ups and potential regimen adjustments are essential based on regular outcome evaluation. By taking into account these multifaceted considerations, the radiation oncologist aims to deliver the most effective, precise, and individualized radiation therapy plan for each patient based on patient details.
    
    \item \textbf{The determination of treatment modality (for instance, proton therapy or photon therapy) contingent on patient details:} Different radiation modality has pros and cons in treating certain disease and usually patients have some unique medical condition, preferences (culture, religion, and/or economics related), and pre-existing co-morbilities. Therefore, it is a complex task for the treating physician to select the most suitable radiotherapy treatment modality for each patient during the initial consultation. Currently, radiation oncologists invest significant hours poring over clinical notes to make an informed decision. The clinical notes are usually very lengthy containing many different text documents including demographics, operative notes, pathology reports, radiology reports, lab results, discharge, and consults, etc. A significant part of clinical notes is prepopulated by subtexts and sessions in the Electronic Medical Record (EMR) system that takes time to read through and digest. This manual process not only consumes time but also introduces the possibility of human error and variability. By analyzing patient data, clinical notes, and previous case studies, LLMs can provide recommendations for the most suitable treatment modality for a specific patient. This could include different types of radiation therapy such as proton therapy, photon therapy, electron therapy, brachytherapy, and their combination, etc. The use of LLMs in this way could help streamline the decision-making process, reduce human error, and potentially lead to better patient outcomes.
    
    \item \textbf{The provision of a Diagnosis description or ICD code in accordance with patient details:} ICD is designed to promote international comparability in the collection, processing, classification, and presentation of mortality statistics. A primary user of ICD codes includes health care personnel, such as physicians and nurses, as well as medical coders, who assign ICD codes to verbatim or abstracted diagnosis or procedure information. LLMs can potentially assist in providing a diagnosis description or ICD code in accordance with patient details. By analyzing patient symptoms, medical history, and other relevant information, LLMs can suggest the most likely diagnosis and corresponding ICD code. This could help streamline the coding process, reduce errors, and ensure more accurate billing and record-keeping. However, it's crucial to remember that these suggestions made by the LLM would still need to be reviewed and confirmed by a healthcare professional. The use of LLMs in this way could potentially improve efficiency and accuracy in healthcare settings.
    
\end{itemize}
This study conducts instruction fine-tuning on three specific tasks, thereby enabling the LLM based on LLAMA2 to acquire specialized knowledge. Consequently, this approach yields answers that closely approximate outputs in clinics for each task.
\subsubsection{ Instruction Tuning Details}

For these disparate tasks, all training data is randomized to ensure comprehensive learning of radiation oncology knowledge by the model, thereby preventing an overemphasis on certain aspects or neglect of specific issues. These models are trained separately, incorporating Low-Rank Approximations (LoRA). The inspiration to employ LoRA stems from its proven stability and reliability in instruction fine-tuning tasks.

The configurations for the experimental training setup included the following:
\begin{itemize}
    \item Batch size: 128
    \item Learning rate: $3 \times 10^{-4}$
    \item Weight decay: 0.01
\end{itemize}

\begin{flushleft}
LoRA parameters:
\begin{itemize}
    \item \texttt{lora\_r} (rank of low-rank factorization): 8
    \item \texttt{lora\_alpha} (scaling factor for the rank): 16
    \item \texttt{lora\_dropout}: 0.05
\end{itemize}
\end{flushleft}

\begin{figure}
    \centering
    \includegraphics[scale=0.5]{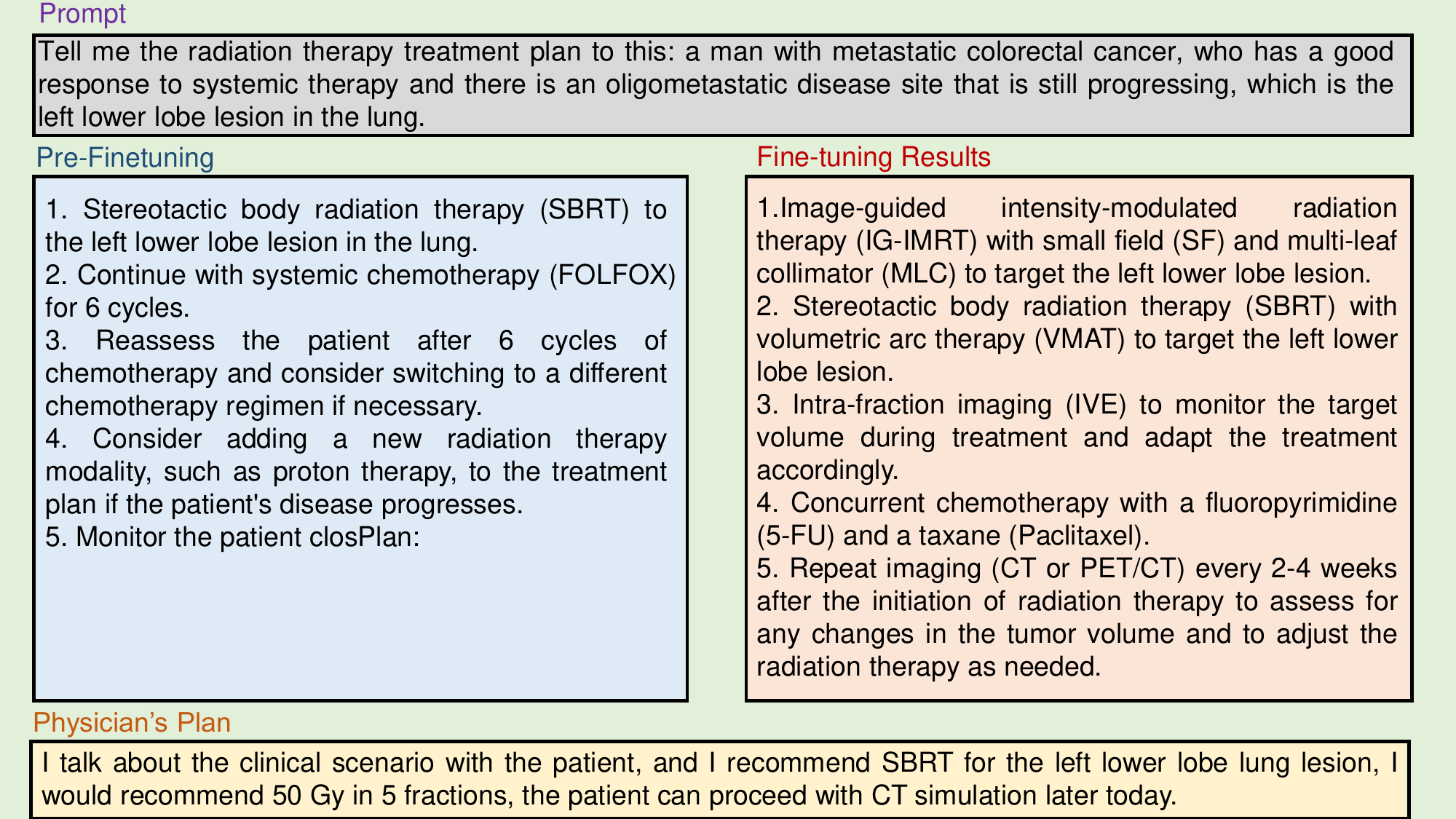}
    \caption{An example of RadOnc-GPT.}
    \label{fig:framework}
\end{figure}

\begin{table}[ht]
\centering
\renewcommand{\arraystretch}{1.2} 
\begin{tabular}{lcccccc}
\toprule
& \multicolumn{3}{c}{RadOnc-GPT} & \multicolumn{3}{c}{Llama2} \\
\cmidrule(lr){2-4} \cmidrule(lr){5-7}
& Rouge-1 & Rouge-2 & Rouge-L & Rouge-1 & Rouge-2 & Rouge-L \\
\midrule
Radiotherapy plan    & 0.4341 & 0.2250 & 0.4271 & 0.0739  & 0.0049 & 0.0657 \\
Treatment modality   & 0.7903 & 0\footnotemark[1]  & 0.7903 & 0.0003 & 0\footnotemark[1] & 0.0003 \\
Diagnosis description / ICD code & 0.7050 & 0.6203 & 0.7026 & 0.0786 & 0.0110 & 0.0609 \\
\bottomrule
\end{tabular}
\caption{Rouge scores of RadOnc-GPT and Llama2. (Rouge score is called as the Recall-Oriented Understudy for Gisting Evaluation (ROUGE) scoring algorithm\cite{lin2004rouge} calculates the similarity between a candidate document and a collection of reference documents.)}
\label{tab:rouge_scores_all}
\end{table}

\footnotetext[1]{"The answer has only one word which means Rouge-2 has no meaning."}


\section{Results}

\subsection{Generation of Radiotherapy Treatment Regimens}

An evaluation of RadOnc-GPT and Llama2 models using Rouge scores was conducted, focusing on their ability to generate radiotherapy treatment regimens. RadOnc-GPT demonstrated a significant edge. Its Rouge-1 score of 0.4341 dramatically surpassed Llama2's 0.0739. The Rouge-2 score for RadOnc-GPT was 0.2250, in stark contrast to Llama2's 0.0049. RadOnc-GPT's dominance persisted in Rouge-L scores, with a score of 0.4271 against Llama2's 0.0657.

The accurate generation of radiotherapy regimens is crucial in radiation oncology. An effective treatment regimen determines a combination of treatment procedures including the feasibility of surgical intervention, the suitability of combining radiation therapy with other treatment modalities such as chemotherapy, and the optimal treatment procedures sequencing to maximize therapeutic outcomes while minimizing toxicity. These decisions are shaped by a holistic evaluation of patient-specific factors, including tumor location and stage, overall health status, and the likely effectiveness of various treatment options in achieving the desired clinical outcomes. RadOnc-GPT's demonstrated prowess here can aid clinicians in formulating optimized treatment strategies, potentially improving patient outcomes and minimizing side effects.

\subsection{Radiation Modality Selection}

In the domain of treatment modality selection, RadOnc-GPT's excellence was even more evident. Both Rouge-1 and Rouge-L metrics stood at an impressive 0.7903, whereas Llama2 lagged significantly with scores of 0.0003. Notably, the Rouge-2 scores were not applicable due to single-word answers.

Treatment modalities in radiation oncology can range from external beam radiation to brachytherapy or even proton therapy. Choosing the correct modality is vital for treatment efficacy and patient safety. RadOnc-GPT's capability to recommend the most appropriate modality, with such accuracy, means treatments could be better tailored to each patient's specific condition, enhancing therapeutic outcomes.

\subsection{Diagnostic Description and ICD Code Prediction}

For diagnostic description and ICD code predictions, RadOnc-GPT's dominance persisted. It achieved a Rouge-1 score of 0.7050, outshining Llama2's 0.0786. RadOnc-GPT secured a score of 0.6203 for Rouge-2, overshadowing Llama2's 0.0110, and a Rouge-L score of 0.7026 against Llama2's 0.0609.

Correctly predicting a patient's diagnosis and aligning it with the proper ICD code is paramount in radiation oncology. It ensures that patients receive the right treatment for their specific cancer type and stage. It also plays a pivotal role in medical record keeping, insurance claims, and research. RadOnc-GPT's demonstrated accuracy in this area can streamline administrative processes and enhance patient care.




The results underscore RadOnc-GPT's superior capabilities in radiation oncology tasks compared to general LLMs. It showed consistent excellence in generating treatment regimens, recommending treatment modalities, and predicting diagnoses description/ICD code. Such an advanced tool can revolutionize the field of radiation oncology. By assisting in generating more accurate treatment regimens and diagnoses description/ICD codes, RadOnc-GPT holds the potential to enhance patient care quality, reduce treatment times, and mitigate complications. 


\section{Discussion}

\subsection{LLMs has huge potentials in radiation oncology}
Transformer architecture on NLP has transformative impact, which has huge potential applications in radiation oncology. The evolution of sophisticated language models like BERT, GPT series, PaLM 2, BLOOMZ, and the Llama series, which have shown diverse capabilities in NLP tasks and downstream applications. In the current research environment, the application of LLMs in the field of radiation oncology remains a relatively unexplored area. Firstly, the confidentiality of patient data is a significant consideration, posing challenges for existing LLMs in learning within this specific domain. However, the field of radiation oncology is highly suitable for LLM training, given the vast accumulation of clinical notes in this sector. If these data can be effectively utilized, we may be able to provide the market with a multifunctional tool with robust clinical utility. This could contribute to the advancement of the field of radiation oncology, enhance treatment accuracy and efficiency, and offer patients higher quality medical services.

\subsection{LLMs can enhance the clinical efficiency in radiation oncology}
LLMs has huge potentials in enhancing patient care and clinical efficiency in radiation oncology. ChatGPT(GPT) already shows the superior performance in radiation oncology physics exams compared to other models and even expert medical physicists. Furthermore, LLMs can leverage existing knowledge to further guide the design of radiation therapy regimens. For instance, they can guide the angular configuration and constraint settings in treatment regimens based on varying radiation therapy techniques and dose prescriptions. These models also hold potential for application in adaptive radiation therapy for treatment replanning \cite{shen2023comprehensive, zhang2018impact, liu2023integrated, yan1997adaptive, yang2021exploratory}. On the basis of this kind of jobs, if researchers can provide a model which are specifically trained on radiation oncology dataset, which definitely will provide new ideas and inject new vitality into the entire clinical radiation oncology community. 

\subsection{LLMs can play an crucial role in connecting data to clinical applications in radiation oncology}
LLMs have a central role in connecting data to clinical applications in radiation oncology. However, this task relies on synthesizing information from various sources, including clinical notes, medical imaging, and dosimetric data. The unstructured nature of clinical notes poses a significant challenge, requiring clinicians to spend extensive hours extracting relevant information. The frequent inclusion of subtexts and templated information from EMR systems makes these notes challenging to navigate. The manual nature of this task not only demands substantial effort but also leaves room for human errors and inconsistencies in decision-making. However, the use of advanced LLMs, specifically fine-tuned on radiation oncology datasets, can mitigate this problem. These models have the capability to quickly sift through complex clinical narratives and suggest the most appropriate radiation treatment, thereby improving efficiency and increasing the chances of achieving the best patient outcomes.

\subsection{LLMs can help radiation oncologists automatically generate insurance claims or appeal documents.}
In the United States, a significant amount of time and resources are expended annually by radiation oncologists and researchers in the generation of insurance claims and appeals. Similarly, insurance companies allocate substantial manpower to process these often monotonous and repetitive textual tasks. The application of LLMs presents a viable solution to streamline these processes for several reasons. Firstly, the industry possesses an extensive corpus of insurance declaration texts, providing a rich dataset for LLMs to train and learn from. Secondly, while these tasks are necessary and often laborious, they are not inherently irreplaceable by non-human resources. LLMs, given systematic and comprehensive training, are fully equipped to undertake these responsibilities, potentially enhancing efficiency and reducing human error. Hence, scholars in the scientific community are encouraged to broaden their perspectives, incorporate LLMs into their practical operations, and enhance productivity across all facets of their work.

\subsection{Limitations}
First, the clinical significance of the outputs produced by our model necessitates additional scrutiny. Although our model surpasses Llama2 in producing contents that are more pertinent, it demands thorough validation to establish its clinical relevance. 

We want to make it clear that our model has been trained exclusively on three tasks: generation of radiotherapy treatment regimens, radiation modality selection (i.e., proton vs. photon), and diagnostic description/ICD code prediction. It does not exhibit generalization capabilities beyond these specified tasks. 

In addition, the evaluation of the generated contents for these tasks relies solely on the ROUGE score, a string-matching based metric that assesses superficial text similarity. However, this metric does not necessarily reflect the depth of semantic meaning, clinical factuality, or relevance.

\section{Conclusion}

In conclusion, this study presents RadOnc-GPT, a pioneering LLM tailored for radiation oncology through rigorous data curation and advanced instruction tuning. Evaluations reveal RadOnc-GPT's superior performance compared to general LLMs in generating clinically coherent and relevant radiation oncology impressions. The model's specialized knowledge and fluency in key radiation oncology tasks underlines the transformative potential of domain-specific LLMs designed with precision. RadOnc-GPT has the potential to significantly improve the speed, accuracy, and quality of radiation therapy decision-making, ultimately benefiting both medical practitioners and the patients they serve. As one of the first investigations applying LLMs to revolutionize radiation oncology, RadOnc-GPT paves the path for specialized medical artificial generative intelligence (AGI) models that can integrate seamlessly into clinical workflows. With diligent oversight, such models can become invaluable assistants to clinicians, unlocking new frontiers in precision radiation oncology. This study serves as a framework to spur future advancements at the intersection of medical AGI and specialized clinical domains.

However, the clinical relevance of our model's outputs is unconfirmed and requires further validation, despite it outperforms Llama2 in generating more relevant contents. The model is fine tuned solely for three specific tasks: radiotherapy regimen generation, modality selection, and diagnostic description/ICD code prediction. And it lacks broader generalization. Moreover, its content evaluation is based only on ROUGE scores, which may not accurately measure semantic depth or clinical accuracy. We will aim to address these limitations in the future work. 

\section*{Acknowledgements}

The authors wish to thank Hongying Feng, Yuzhen Ding, Haixing Dai, Zihao Wu, and Lin Zhao for their invaluable feedback and insights during the development and evaluation of RadOnc-GPT.

\bibliography{RadOnc-GPT_refs}
\bibliographystyle{unsrt}

\end{document}